\documentclass[twocolumn,%
 reprint,
 amsmath,amssymb,
 aps,
prl,
]{revtex4-1}

\usepackage[usestackEOL]{stackengine}
\usepackage{graphicx}
\usepackage{dcolumn}
\usepackage{bm}

\usepackage{enumerate}
\usepackage{nicefrac}
\usepackage{lipsum}

\newcommand{\bea}{\begin{eqnarray}}
\newcommand{\eea}{\end{eqnarray}}

\usepackage{centernot}
\usepackage{color}
\newcommand{\be}{\begin{equation}}  
\newcommand{\ee}{\end{equation}} 

\makeatletter
\let\Hy@backout\@gobble
\makeatother

\begin{document}


\title{Quantum Forces from Dark Matter and Where to Find Them}

\author{Sylvain Fichet}
\email{sylvain@ift.unesp.br}
\affiliation{%
ICTP-SAIFR \& IFT-UNESP, R. Dr. Bento Teobaldo Ferraz 271, S\~ao Paulo, Brazil
}

\begin{abstract}

 We observe that sub-GeV Dark Matter (DM) induces Casimir-Polder forces between nucleons, that can be accessed by experiments from  nuclear to molecular scales.   
 We calculate the nucleon-nucleon potentials arising in the DM effective theory and note that their main features are fixed by dimensional analysis and the optical theorem. 
 Molecular spectroscopy and neutron scattering turn out be  DM search experiments, and are found to be  complementary to nucleon-based DM direct detection.
  Existing data  set limits on  DM with mass up to $\sim 3-50$ MeV and with effective interaction up to the $O(10-100)$ MeV scale, constraining a region typically difficult to reach for other experiments.



\end{abstract}

\pacs{Valid PACS appear here}

\maketitle

\section{Introduction}
  
A  body of evidence suggests that our Universe is filled with an unknown Dark Matter (DM), which may be  a new kind of particle lying beyond  the Standard Model (SM) of particle physics. What do we know about this putative dark particle? Apart from its weak interaction with photons, very little is known about its properties, including mass, spin and couplings. Importantly, a robust lower bound  is set from structure formation in the Universe: Galaxy formation implies the dark particle  mass should satisfy $m\gtrsim 2$\,keV \cite{Bond:1983hb,Viel:2013apy,Menci:2016eui} to limit its free-streaming length.  



Although direct detection experiments have reached impressive sensitivity above the GeV mass scale, 
 the dark particle remains so far elusive \cite{Undagoitia:2015gya}. 
For masses below the GeV,  direct detection methods lose sensitivity because nuclear recoil becomes too soft to be detected. At the LHC, monojet searches could be sensitive to a dark particle below the GeV if it has a contact interaction with the SM particles. However when the interaction between the DM and SM  is resolved
at the LHC,  the sensitivity is expected to vanish when the scale of this ``portal'' becomes too low. For instance, for mediation via a $Z'$ particle, the sensitivity vanishes   below roughly $O(10-100)\,$GeV \cite{Khachatryan:2014rra}.

As the two most direct search techniques---scattering on nucleons and monojets---are inefficient when DM is sub-GeV and the portal scale is light (forming thus  a ``light dark sector''), other experiments need to be devised. Cosmological and astrophysical constraints can of course play a role but are somewhat indirect an depend on many assumptions, hence more direct searches for light dark particles are certainly needed. As a matter of fact, an increasing number of ideas are being proposed to search for sub-GeV DM, including  semiconductor\,\cite{Essig:2011nj,Graham:2012su, Essig:2015cda,Lee:2015qva}, superconductor\,\cite{Hochberg:2015pha,Hochberg:2015fth}, and superfluid\,\cite{Guo:2013dt,Schutz:2016tid} targets, carbon structures \cite{Cavoto:2016lqo,Hochberg:2016ntt}, crystals\,\cite{Essig:2016crl,Budnik:2017sbu}, scintillators\,\cite{Derenzo:2016fse}, electron scattering or Bremsstrahlung in conventional detectors\,\cite{Essig:2012yx, Essig:2017kqs, Kouvaris:2016afs}, neutrino fixed target experiments\,\cite{Batell:2009di,deNiverville:2011it,deNiverville:2012ij,Dharmapalan:2012xp,Batell:2014yra,Soper:2014ska,Dobrescu:2014ita,Frugiuele:2017zvx}
 and the SHiP proposal\,\cite{Alekhin:2015byh}. All these proposals rely on dark particles on the mass-shell.

 In this Letter, we take a different approach, by considering a phenomenon induced by \textit{virtual} dark particles (see also \cite{Fichet:2016clq}). 
 We point out that whenever sub-GeV Dark Matter couples to nucleons, it induces Casimir-Polder  forces between them.
 The $m\gtrsim 2$\,keV  bound from structure formation  corresponds then to a maximum   scale of $\sim 1$\,\AA~for the force range,  implying that these DM-induced forces can be active up to molecular scales. 
We will see that molecular and atomic precision spectroscopy as well as neutron scattering experiments are sensitive to such forces. Existing data from these research fields will be exploited in this Letter to obtain new limits on sub-GeV DM. In the following we will refer to the DM-induced Casimir-Polder forces  simply as \textit{DM forces}.

\begin{figure}[t]
  \centering
  \includegraphics[scale = 0.6,trim={0 0.cm 0 0},clip]{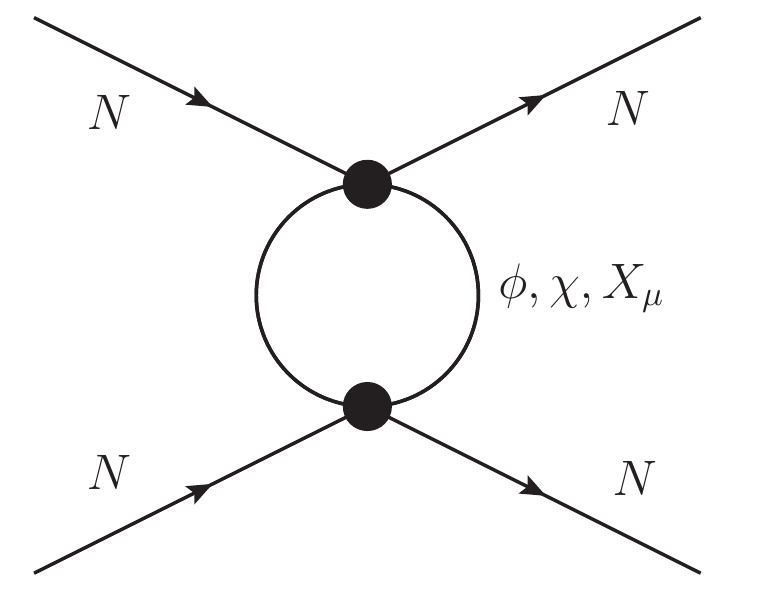}
   \caption{ 
The exchange of two dark particles  inducing a force between the nucleons.  }
  \label{fig:4fermion}
\end{figure}

The focus in this work is on DM which interacts with nucleons. 
Interactions with electrons could also be studied although they are already constrained by $e^+e^-$ collisions\,\cite{Hewett:2012ns}. The DM forces between electrons and nucleons could be studied by analyzing King plots from isotope shift spectroscopy, a technique recently proposed in \cite{Frugiuele:2016rii,Delaunay:2016brc,Delaunay:2016zmu,Berengut:2017zuo}.

Our approach relies on virtual dark particles, hence all our results apply whether the dark particle is stable or not--an agnostic viewpoint is taken in the companion paper Ref.~\cite{us}. Yet, the implications  for Dark Matter deserve special attention given the profusion  of experimental and theoretical  activities in this area. This Letter is thus focused on the implications for DM searches and for some predictive DM cosmological scenarios.

%

%



 \section{Casimir-Polder forces from Dark Matter effective theory}

Whenever DM interacts with light quarks or with gluons, it couples to nucleons below the QCD scale. We use an effective field theory  (EFT) to describe the DM interactions with  nucleons. Our most general results for DM forces are given here, but more details on EFT and calculation are available in Ref.~\cite{us}.

   In the limit of unpolarized nonrelativistic nucleons, only the interactions involving $\bar N N,\bar N \gamma^\mu  N$ are relevant. The DM particle is noted $\phi$, $\chi$ and $X$ for particles of spin $0$, $1/2$, $1$  respectively - either self-conjugate or not. Results will be presented for a representative subset of interactions for DM particles of each spin,
\small
\begin{align}  \label{eq:Leff}
{\cal O}^0_a&=\frac{1}{\,\Lambda} \bar N N |\phi|^2  \,,\quad {\cal O}^{\nicefrac{1}{2}}_a=\frac{1}{\,\Lambda^2} \bar N N \bar \chi \chi \,, \nonumber
\\ \nonumber
{\cal O}^0_b&=  \frac{1}{\,\Lambda^2}\, \bar N \gamma^\mu N \phi^* i\overleftrightarrow\partial_\mu  \phi
  \,,\quad  {\cal O}^{\nicefrac{1}{2}}_b=\frac{1}{\,\Lambda^2} \bar N \gamma^\mu N \bar \chi \gamma^\mu \chi \,,
\\ \nonumber
{\cal O}^0_c&=\frac{1}{\,\Lambda^3} \bar N N \partial^\mu \phi^* \partial_\mu\phi\,,\quad
{\cal O}^{\nicefrac{1}{2}}_c=\frac{1}{\,\Lambda^2} \bar N \gamma^\mu  N \bar \chi \gamma^\mu \gamma^5\chi \,,
     \\  \nonumber
{\cal O}^{1}_a&=\frac{m^2}{\Lambda^3} \bar N N |X^\mu+\partial^\mu \pi|^2  \,, \\ \nonumber
{\cal O}^{1}_b&=\frac{1}{\Lambda^2} \,2 \bar N \gamma^\mu N \,{\rm Im}(X^*_{\mu\nu }X^\nu+ \partial^\nu(X_\nu X^*_\mu) +\partial^\mu \bar c c^*) \,, \\
{\cal O}^{1}_c&=\frac{1}{\Lambda^3} \bar N N |X^{\mu\nu}|^2 \,, \quad {\cal O}^{1}_d=\frac{1}{\Lambda^3} \bar N N X^{\mu\nu}\tilde X^{\mu\nu}\,.
\end{align}
\normalsize
where $\overleftrightarrow\partial=\overrightarrow\partial-\overleftarrow\partial$, and $2 {\rm Im}(X^*_{\mu\nu }X^\nu)+\ldots$ corresponds to the gauge current for $X$. $\pi$ and $c, \bar c$ are respectively the Goldstone and ghosts accompanying $X$. 
  The ${\cal O}^s_b$ operators involve gauge currents of DM, which  vanish if DM is self-conjugate (real scalar or vector, Majorana fermion). For simplicity we will assume a universal coupling to protons and neutrons---all our results are easily generalized for non-universal couplings to nucleons. 
  

The effective interactions of Eq.~\eqref{eq:Leff} induce 4-nucleon interactions when closing the DM loop,   suggesting to search for DM in nucleon interactions.
The diagram  being a loop amplitude in the EFT,  local 4-nucleon interactions are in principle also present \cite{Manohar:1996cq}, that depend on the UV completion of the theory (such as heavy mediators or intrinsic polarizability\,\cite{Fichet:2016clq}). However, the experimental results we will use are by desgin fully or nearly independent of these local terms and are thus perfectly appropriate to specifically target a light dark particle. 



The  force between nucleons induced by the diagram in Fig.\,\ref{fig:4fermion}  is obtained by taking the nonrelativistic limit of the amplitude, taking the 3d Fourier transform, continuing in the complex plane, and reduces to integrating over a branch cut.
 The subsequent potentials are given by modified Bessel functions evaluated at $2m r$, $K_i(2m r)\equiv K_i $. The operators of Eq.\,\eqref{eq:Leff} give the DM forces
\small
\begin{align}
\bigg(& V^0_a, V^0_b, V^0_c, V^{\nicefrac{1}{2}}_a, V^{\nicefrac{1}{2}}_b, V^{\nicefrac{1}{2}}_c,  V^1_a, V^1_b, V^1_c, V^1_d\bigg)=  \frac{1}{32 \pi^3\,r }\times \nonumber  \\ \nonumber
 \Bigg( \,&-2^\eta\frac{(C^0_a)^2}{ \Lambda^2 } \frac{m}{r} K_1\,, 
 \eta\frac{4\,(C^{0}_b)^2}{\Lambda^4  } \frac{m^2}{r^2} K_2\,, 
 \\ \nonumber
 &-2^\eta\frac{(C^0_c)^2}{\Lambda^6} \left(\left(\frac{30m^2}{r^4}+\frac{6m^4}{r^2}\right) K_2+ \left(\frac{15m^3}{r^3}+\frac{m^5}{r}\right) K_1 \right), \nonumber 
 \\ \nonumber 
 & -2^\eta\frac{12\,(C^{\nicefrac{1}{2}}_a)^2}{\Lambda^4  } \frac{m^2}{r^2} K_2\,, 
\eta\frac{16 (C^{\nicefrac{1}{2}}_b)^2 \,m^2}{\Lambda^4  } \left(\frac{1}{r^2}K_2+ \frac{m}{r} K_1 \right)\,,
 \\ \nonumber 
& 2^\eta\frac{8\,(C^{\nicefrac{1}{2}}_c)^2}{\Lambda^4  } \frac{m^2}{r^2} K_2\,,  
-2^\eta\frac{(C^1_a)^2}{\Lambda^6} \left(\left(\frac{15m^2}{r^4}+\frac{3m^4}{r^2}\right) K_3 \right)\,,
\\ \nonumber 
& \eta\frac{(C^{1}_b)^2}{\Lambda^4  } \left(\frac{40m^2}{r^2} K_2+\frac{36m^3}{r} K_1\right)\,, \\
&   -2^\eta\frac{12(C^1_c)^2}{\Lambda^6} \left(\left(\frac{20m^2}{r^4}+\frac{4m^4}{r^2}\right) K_2+ \left(\frac{10m^3}{r^3}+\frac{m^4}{r^4}\right) K_1 \right), \nonumber
 \\
&   -2^\eta\frac{24(C^1_d)^2}{\Lambda^6} \left(\frac{2m^4}{r^2}K_3+\frac{5m^3}{r^3}K_2\right) 
~~ \Bigg) \,.  \label{eq:forces}
\end{align}
\normalsize
The forces from the ${\cal O}_{a,c}^s$ operators (``scalar channel'') are attractive and those from  the ${\cal O}_{b}^s$ operators (``vector channel'') are repulsive.


The main features of these forces can be understood using both dimensional analysis and the optical theorem applied to the diagram of Fig.\,\ref{fig:4fermion}. First, the optical theorem   dictates the sign of the discontinuity over the branch cut, and thus the sign of the potentials.   Second, the short-distance behavior  
$(r^{-3},r^{-5},r^{-7},r^{-5},r^{-5},r^{-5},r^{-7},r^{-5},r^{-7},r^{-7})$ of Eq.~\eqref{eq:forces} is dictated by dimensional analysis. Third, using the optical theorem, the long-distance behavior $e^{-2 m r}(r^{-\frac{5}{2}},r^{-\frac{7}{2}},r^{-\frac{5}{2}},r^{-\frac{7}{2}},r^{-\frac{5}{2}},r^{-\frac{7}{2}},r^{-\frac{5}{2}},r^{-\frac{5}{2}},r^{-\frac{5}{2}},r^{-\frac{7}{2}})$ is related to velocity-suppression of the $\bar N N \leftrightarrow \chi \bar \chi$ amplitude at $s\sim 4m^2$  (see details in \cite{us}).

 \section{ Dark Matter bounds from molecular spectroscopy }

Impressive progresses on both experimental
\cite{Niu201444,Biesheuvel:2016azr,UbachsAPB, Balin2011, Hori2011,PhysRevLett.98.173002,PhysRevLett.108.183003,Ubachs09}  and theoretical \cite{Karr14,PhysRevLett.113.023004,Karr2016,KHK14,PhysRevA.76.022106,PhysRevA.82.032509, 0953-4075-37-11-010,PhysRevA.74.052506,PhysRevA.77.042506,PhysRevA.77.022509,doi:10.1021/ct900391p,Pachu11}
 sides of precision molecular spectroscopy  have been accomplished recently, 
opening the possibility of testing new physics below the \AA\,scale using  transition frequencies of well-understood simple molecular systems. Certain of these results have recently been used to bound short distance modifications of gravity, see Refs.\,\cite{PhysRevD.87.112008,Salumbides:2013dua,Ubachs17,Ubachs20161,Ubachs20161}.

The most relevant systems for which both precise measurements and predictions are available are the hydrogen molecule H$_2$, the molecular hydrogen-deuterium ion HD$^+$ and muonic molecular deuterium ion $dd\mu^+$, where $d$ is the deuteron. This last system is exotic in the sense that a heavy particle (the muon) substitutes an electron. As a result the internuclear distance is reduced, providing a sensitivity to  forces of shorter range, and thus to heavier dark particles. 

 The presence of the DM force shifts the energy levels by $\Delta E = \int d^3 {\bf r}\, \Psi^*(r) V(r) \Psi(r)$  at first order in perturbation theory. 
We compute these  energy shifts for the transitions between the  $(\nu=1,J=0)-(\nu=0,J=0) $ states for $H_2$, $(\nu=4,J=3)-(\nu=0,J=2) $ of HD$^+$,  and the binding energy of the $(\nu=1,J=0)$ state of $dd\mu^+$ using the wave functions given in \cite{Salumbides:2013dua,Ubachs17,Vlad}, with $\nu,J$ being respectively the rotational and vibrational quantum numbers. The average internuclear distances for the quantum states considered are respectively $\sim 1$\,\AA\,for H$_2$, HD$^+$,  and $\sim 0.005-0.08$\,\AA\,for $dd\mu^+$.  
Bounds on the DM forces are obtained using  combined uncertainties  of respectively  $3.9$\,neV\,\cite{PhysRevD.87.112008, Ubachs17}, 
$0.33$\,neV\,\cite{PhysRevD.87.112008}, 
$\delta E=0.7$\,meV\,\cite{Salumbides:2013dua}. For each observable the experimental uncertainty is slightly larger than the theoretical one, at most by an order of magnitude\,\footnote{For $dd\mu^+$, the predicted energy shifts from the dark particle can also have some UV sensitivity, see \cite{us}.}.
 Therefore progresses on both experimental and theory sides  would be  needed in order to improve the sensitivity of these molecular observables.

The lower bounds obtained on  $\Lambda$ and $m$ are typically of order $10-100 $\,MeV and  of $3-50 $\,MeV respectively.  It turns out that limits from $dd\mu^+$ are the most stringent on both $\Lambda$ and $m$.
Samples results are shown in Figs.\,\ref{fig:DD},\,\ref{fig:exclusions}.


 \section{ Dark Matter bounds from neutron scattering }

Progresses in measuring the scattering of cold neutrons on nuclei have been recently made and have been used to put bounds on short-distance modified gravity, \cite{Nesvizhevsky:2004qb,Leeb:1992qf,Frank:2003ms,Watson:2004vh,Greene:2006qj, Baessler:2006vm,Nesvizhevsky:2007by,Kamiya:2015eva}. 
The cold neutron scattering cross-section can be measured at zero angle by ``optical'' methods,  at non-zero angles using Bragg diffraction, or over all angles by the ``transmission'' method giving then the total cross-section \cite{Koester199165}. 

In the following we adapt the analyses of \cite{Nesvizhevsky:2007by}  to the DM case.
At low energies the standard neutron-nuclei interaction is a contact one. New physics can induce both contact and noncontact contributions, and it is convenient to introduce the scattering length $ \sqrt{\frac{\sigma({\bf q})}{4\pi}}\equiv l({\bf q})=l_{\rm std}^{\rm C}+l_{\rm NP}^{\rm C}+l_{\rm NP}^{\rm NC}({\bf q}) $, where the $l_{\rm std}^{\rm C}$, $l_{\rm NP}^{\rm C}$  terms are independent of momentum transfer ${\bf q}$. A convenient way to look for an anomalous interaction is to search for 
$l_{\rm NP}^{\rm  NC}({\bf q})$ by comparing the scattering length obtained by different methods, $l_{\rm Bragg}/l_{\rm opt}$, $l_{\rm tot}/l_{\rm opt}$. This approach eliminates $l_{\rm std}^{\rm C}$ but also $l_{\rm NP}^{\rm C}$, and is therefore only sensitive to the nonlocal part of the scattering potential, which corresponds to the DM force.  
Sample results from  $l_{\rm Bragg}/l_{\rm opt}$, $l_{\rm tot}/l_{\rm opt}$ are shown on Figs.~\ref{fig:DD},\,\ref{fig:exclusions}. The best sensitivity comes from $l_{\rm tot}/l_{\rm opt}$, which typically competes with the reach from $dd\mu^+$. 


\section{Complementarity with direct detection}

\begin{figure}
\includegraphics[width=7.5 cm,trim={0 0.1cm 0 0},clip]{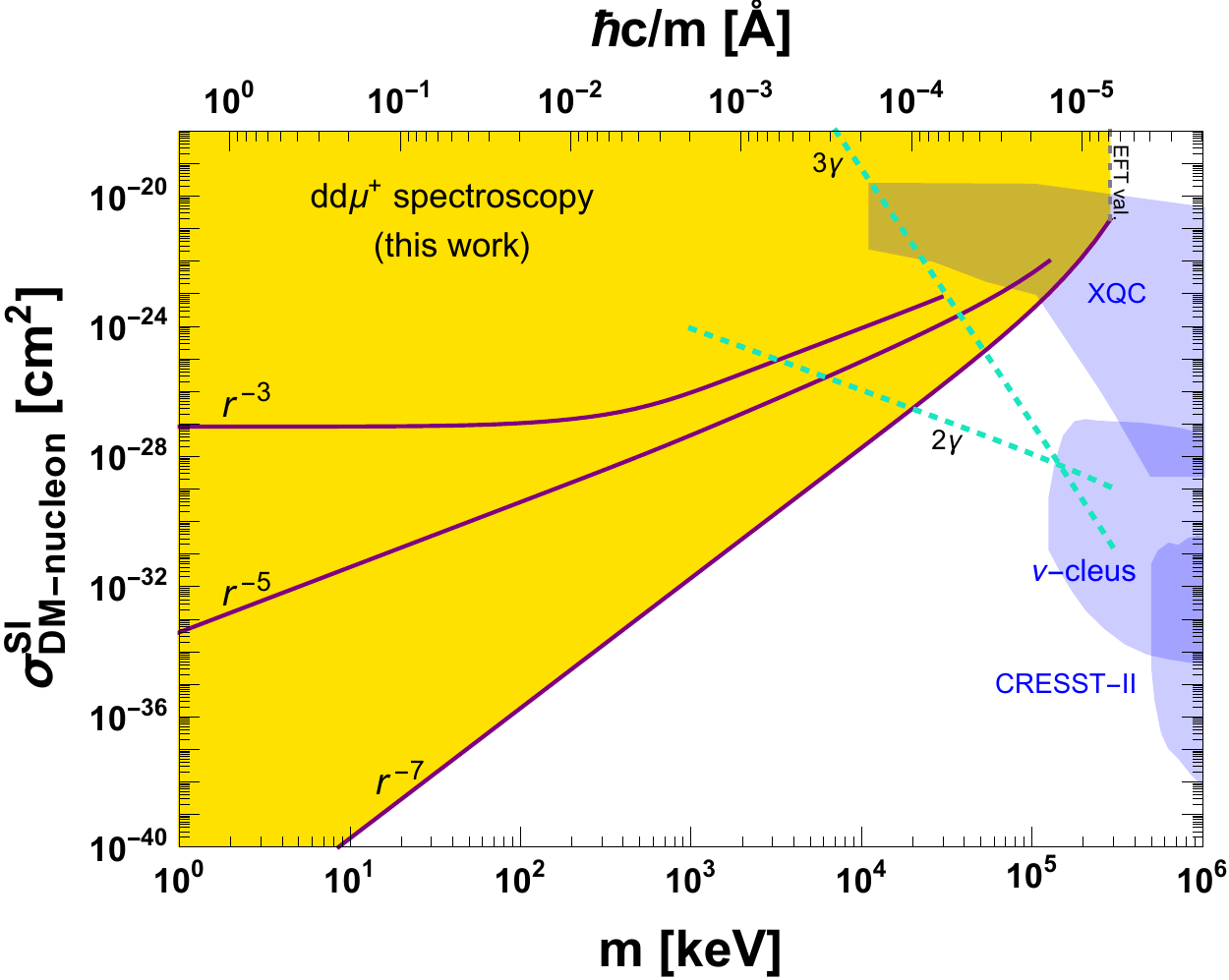}\\
\vspace{0.25cm}
\includegraphics[width=7.5 cm,trim={0 0 0 1.3cm},clip]{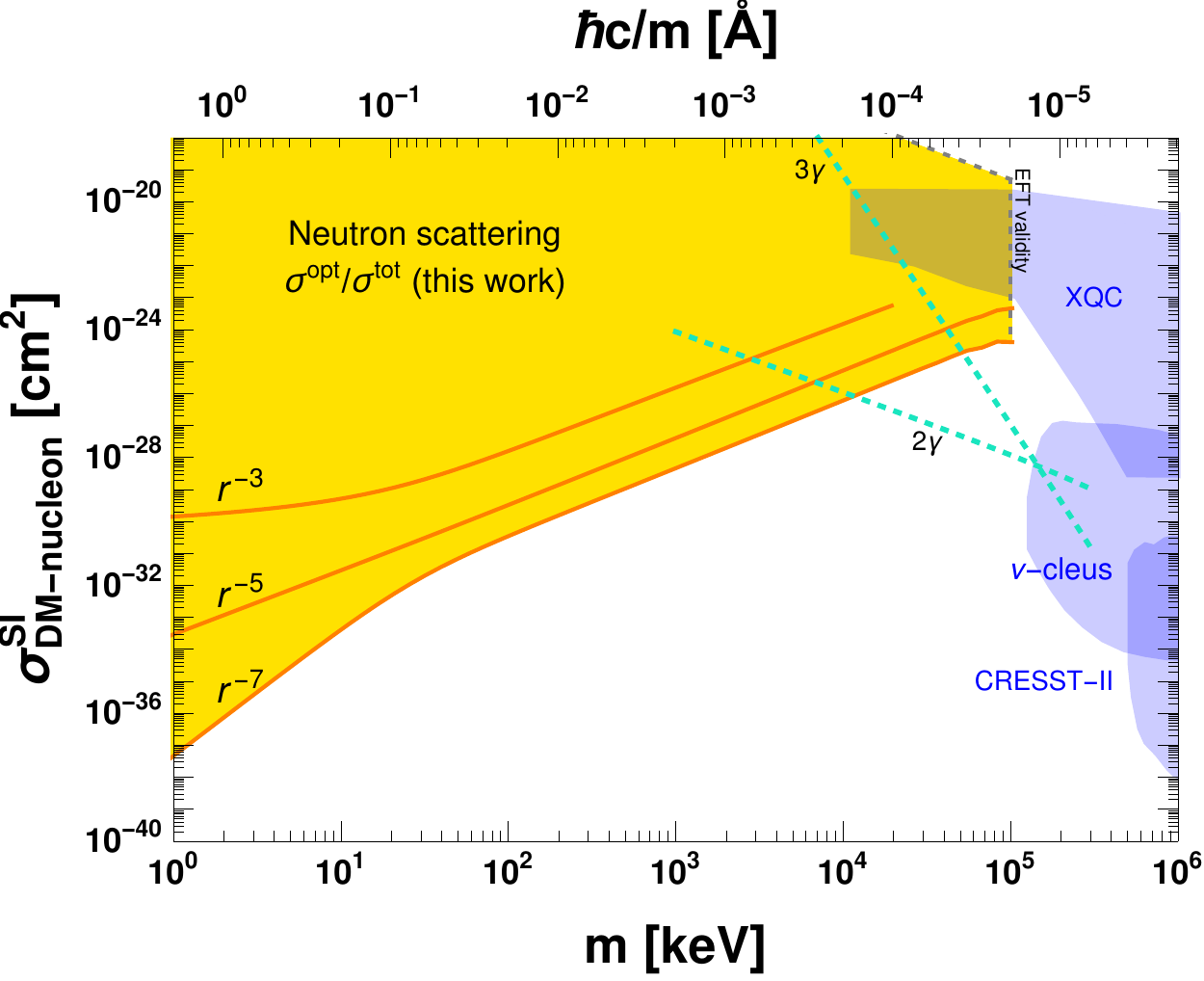}
\caption{Leading exclusion regions from DM forces translated into the $\sigma^{\rm SI}$-$m$ plane of direct detection. Exclusion regions from a sample of forces with $r^{-3}$, $r^{-5}$, $r^{-7}$ short-distance behaviour are shown. Direct detection bounds labelled XQC, $\nu$-cleus, CRESST-II are respectively from Refs.~\cite{Erickcek:2007jv}, \cite{Davis:2017noy}, \cite{Angloher:2015ewa}. Dotted lines correspond to  $\Omega h^2=0.112$ with thermal freeze-out of Dirac DM $\chi \bar \chi \rightarrow 2\gamma, 3\gamma$ pion-driven annihilations. 
\label{fig:DD}
}
\end{figure}

Searches for real DM scattering off nucleons  (\textit{i.e.} DM direct detection) can be described by the same effective operators as the ones used for the DM force, hence we can readily compare both techniques. It is convenient to translate our DM  forces bounds into equivalent exclusion regions on the spin-independent DM-nucleon scattering cross-section $\sigma^{SI}$. The exclusion regions in the 
 $\sigma^{SI}-m$ plane are shown in Fig.~\ref{fig:DD}. The complementarity is clear: the direct detection sensitivity always vanishes for small  $m$, while our bounds from DM forces vanish at large $m$ and are valid  down to zero mass. The typical exclusion regions can be conveniently classified with respect to the short-range behaviour of the DM force. The $r^{-3},r^{-5},r^{-7}$ regions shown in Fig.~\ref{fig:DD} are from ${\cal O}^0_a$, ${\cal O}^{\nicefrac{1}{2}}_b$, ${\cal O }^1_c$. Interestingly, the operators ${\cal O}^{\nicefrac{1}{2}}_c$, ${\cal O}^1_d$ have vanishing SI cross-section, while they are probed by the DM force.

\section{Cosmology}

There is a lot of freedom to accomodate the observed relic abundance of DM with sub-GeV masses, see \textit{e.g.}  \cite{Batell:2014yra,Dobrescu:2014ita,Kuflik:2015isi,Bernal:2015ova,Bernal:2015xba}, but in general such scenarios are independent of the DM-nucleon coupling.  
Here we rather present two versions of a scenario which relies solely on the coupling to hadrons to explain the  DM  abundance, and can be thus meaningfully confronted to our DM force bounds---and to  direct detection.


\textit{Loop-level freeze-out}
As DM interacts with nucleons via the operators of Eq.~\eqref{eq:Leff}, we can assume it couples similarly to other hadrons. Whenever DM interacts with charged hadrons,  DM annihilation into photons can always occur by closing the hadron loop, and thermal freeze-out is controlled by this annihilation.  Annihilation is  mostly into 2$\gamma$ for the scalar channel, but only into 3$\gamma$ for the vector channel. Main contribution is from charged pions.   
Focussing on Dirac DM annihilating as $\chi \bar\chi\rightarrow \gamma\gamma$ via ${\cal O}_a^{\nicefrac{1}{2}}$ and in  $\chi \bar\chi\rightarrow \gamma\gamma\gamma$ via  ${\cal O}_b^{\nicefrac{1}{2}}$ (possibly UV-completed by a $Z'$),  taking the heavy pion limit and deducing the local $\gamma\gamma\gamma Z'$ vertex  from \cite{Baldenegro:2017aen}, we get order-of-magnitude estimates (assuming same coupling to pion and nucleons)  \small
\be
\left<\sigma v\right>_{ 2 \gamma}\sim \left(\frac{2 \cdot 10^{-5}}{ {\rm GeV^2}} \right) \frac{m^4}{\Lambda^4} \,, \quad \left<\sigma v\right>_{ 3 \gamma}\sim \left( \frac{ 0.1 }{{\rm GeV}^8} \right) \frac{m^{10}}{\Lambda^4} \,.
\ee
\normalsize
This minimal scenario is shown in Figs.~\ref{fig:DD},\, \ref{fig:exclusions}.   

\textit{Phase-transition-induced freeze-out} 
It is also possible that the DM interactions with hadrons take the form  Eq.~\eqref{eq:Leff} only below a phase transition at  $T\equiv f$. This happens in particular if the mediator   gets a mass after phase transition, \textit{e.g.} a scalar or a $Z'$ getting mass via a dark Higgs mechanism. In such scenario, the decoupling  occurs from the symmetry breaking instead of the expansion of the Universe, and the DM relic abundance depends \textit{only} on $m$ and $f$. In the case of mediation by a massless species before transition, the parameter space is simply given by $\Lambda>\Lambda_{\rm min}$, where $\Lambda_{\rm min}$ is the value required for thermal freeze-out occurring in the previous scenario. We notice another intriguing realization of this mechanism: DM could actually \textit{appear} at the phase transition as a result of the confinement of a strongly-interacting gauge theory. It is plausible that DM Boltzmann suppression occur during the phase transition, because chemical equilibrium should be conserved to some extent, at least for a crossover. At large $N$ and t'Hooft coupling this scenario admits a  holographic description and could be studied in this fashion.

In all the above scenarios, the annihilation into photons, if active below neutrino decoupling ($T\sim 2.3\, {\rm MeV}$), reheats photons and tends to reduce the observed effective number of neutrinos ($N_{\rm eff}$) \cite{Boehm:2013jpa,Nollett:2013pwa,Nollett:2014lwa}. However any extra relativistic species - just like the light mediators present in the second mechanism - can increase $N_{\rm eff}$ back. 
For both scenarios, we require the freeze-out temperature to be above $0.1$ MeV to avoid changes in the He abundance from Big Bang Nucleosynthesis \cite{Pospelov:2010hj}, hence $m\gtrsim 1 \, {\rm MeV}$. DM annihilation into $2\gamma$ produces $\gamma$-ray lines but is velocity-suppressed  for Dirac DM.

 \section{ Complementarity with  other experiments }

 \begin{figure}
\begin{picture}(250,245)
\put(0,0){\includegraphics[width=8 cm,trim={0 0 0 0},clip]{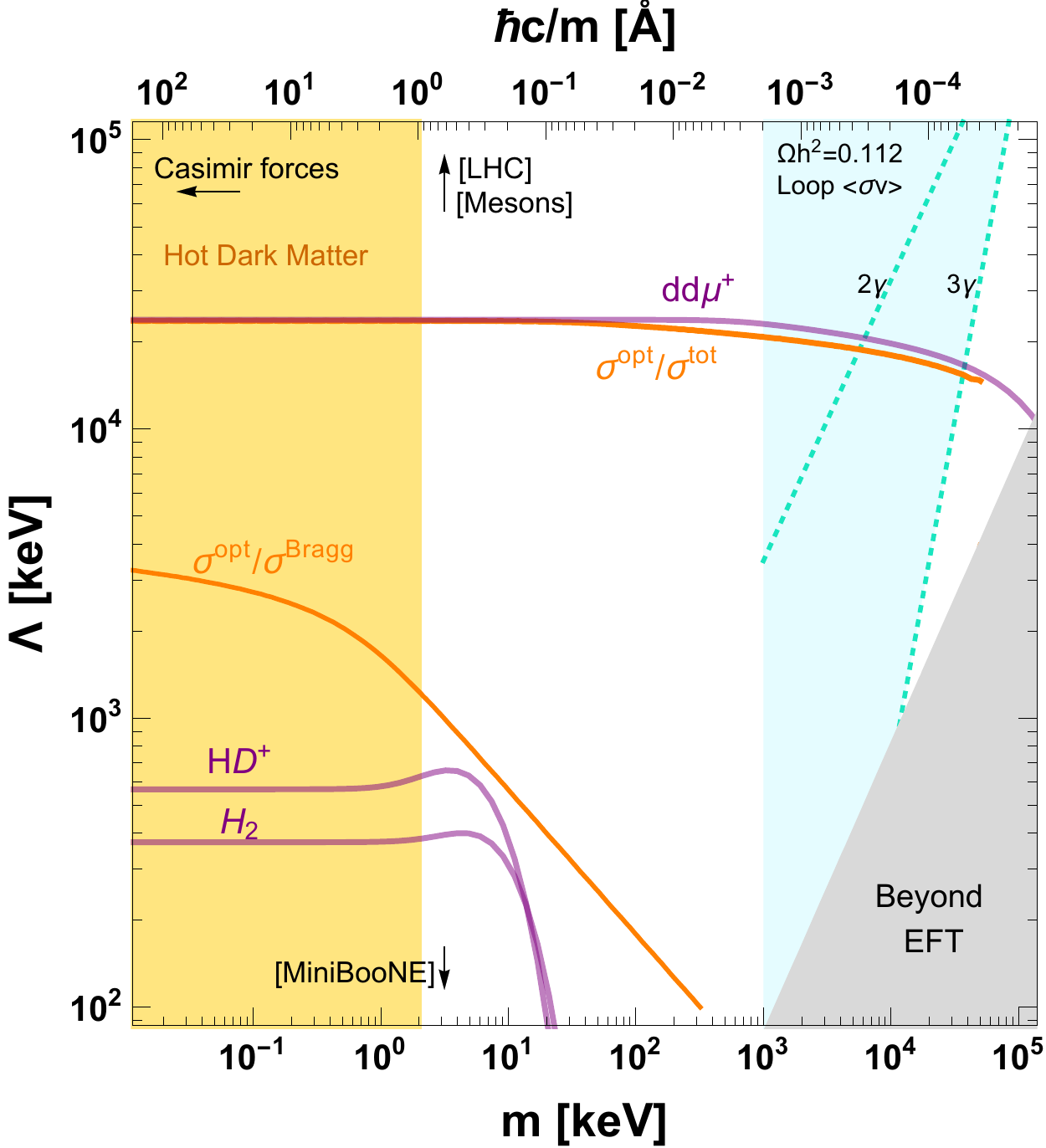}}
\end{picture}
\caption{
Limits on the ${\cal O}^{\nicefrac{1}{2}}_b$ interaction of Dirac DM in the $m-\Lambda$ plane. Regions excluded by precision molecular spectroscopy and  neutron scattering are show by purple and red upper boundaries respectively. The constraint from structure formation \cite{Bond:1983hb,Viel:2013apy,Menci:2016eui} is shown in orange. Exclusion regions lying outside the shown parameter space (see text) are indicated with arrows. Those in brackets  assume the interaction is mediated by a light leptophobic $Z'$ with $m_{Z'}\lesssim 10$\,MeV, $\alpha_B\lesssim 0.03 \alpha_{\rm em}$. We indicate the exclusion regions from Casimir force searches \cite{Lamoreaux:1996wh,Jaeckel:2010ni}, from LHC missing energy searches \cite{Khachatryan:2014rra}, from kaon and quarkonium invisible decays \cite{Artamonov:2009sz,Tajima:2006nc,Aubert:2009ae,Ablikim:2007ek} assuming flavour universal couplings, and the sensitivity region from the MiniBoone detector\,\cite{Dharmapalan:2012xp,Batell:2014yra,Frugiuele:2017zvx}.  In the gray region the UV completion (\textit{e.g.} $Z'$ exchange) has to be specified.  
Dotted lines correspond to  $\Omega h^2=0.112$ for thermal freeze-out with Dirac DM $\chi \bar \chi \rightarrow 2\gamma, 3\gamma$ annihilations. Blue regions correspond to phase-transition-induced  freeze-out. 
\label{fig:exclusions}
}
\end{figure}

Let us finally compare the constraints from DM forces to other existing ones -apart from direct detection.  From the basic assumption that DM couples to nucleons, there are constraints from Casimir force measurements and pendulum experiments, starting at $m\sim 10-100$\,eV. However apart from scalar DM  with the  ${\cal O}_{a}^0$ operator, all the other forces are best constrained at any mass by the methods presented here  \cite{us}. 

 For particle physics experiments, our low-energy  EFT breaks down and comparison has to be on the basis of a specific UV completion for a given DM candidate. We focus on Dirac DM with ${\cal O}^{\nicefrac{1}{2}}_b$ effective interaction,  which is naturally UV-completed by a $Z'$ boson from a hidden $U(1)$ coupling to quark and $\chi$ currents with strength $g_B$. At low-energy the mapping  onto ${\cal O}^{\nicefrac{1}{2}}_b$ is given by $\Lambda^{-2}=-g_B^2 m_{Z'}^{-2}$. The  $Z'$ is  leptophobic and with no kinetic mixing to the photon.
 We will translate  constraints on the $Z'$ model into bounds on the $\Lambda$ parameter. An important subtlety  is that these translated constraints \textit{do not} necessarily take the form of a lower bound on $\Lambda$, because they originate from a UV completion.

 Constraint from kaon  decay $K^+\rightarrow \pi^++ {\rm invisible} $ using the  bound from \cite{Artamonov:2009sz} and the prediction from \cite{Batell:2014yra} (see Eq.~(7)) gives $g^{\,}_B m^{\,}_{Z'}<0.13$\,MeV.  
   In the low-energy EFT  this becomes $ \Lambda\lesssim 0.01\,{\rm MeV}\alpha_B^{-1} $.  This is an \textit{upper bound} on $\Lambda$,  and there is thus complementary  with the DM force, which instead sets a lower bound.  A similar, subleading bound comes from  $J/\Psi$ invisible decay (\cite{Ablikim:2007ek,Dobrescu:2014ita, Fernandez:2014eja}).
     The bound from kaon decay would reach down $\Lambda\sim10$\,MeV and start to compete with the DM force for $\alpha_B\gtrsim 0.1 \alpha_{\rm em}$. 
Monojet with missing energy (see e.g. \cite{Khachatryan:2014rra}) and dijet searches from the LHC are also constraining the $Z'$ scenario. The sensitivity drops when the $Z'$ gets light, hence these searches put upper bounds on $\Lambda$ which however do not compete with the meson bounds.  
Finally, following Ref.\,\cite{Batell:2014yra,Frugiuele:2017zvx}, measurements at the MiniBooNE detector \cite{Dharmapalan:2012xp} 
are expected to provide an upper bound on $g_B$ for values of $m_{Z'}$ down to $\sim 10-100 $\,MeV, which implies a  lower bound on $\Lambda$. For lighter $Z'$, the sensitivity should drop, implying that the region constrained by the MiniBooNE measurements lies at smaller $\Lambda$, as indicated in Fig.\,\ref{fig:exclusions}. 

In this $Z'$ scenario,  particle physics experiment do not access the  region with roughly $m_{Z'}<10$\,MeV and $\alpha_B\lesssim 0.1 \alpha_{\rm em}$. The DM force measurements are probing part of this region, and are thus complementary to the other experiments.

\section{Summary}

We have calculated the quantum forces induced by sub-GeV DM coupled to nucleons, and  we have shown that molecular spectroscopy and neutron scattering can be used as DM search experiments.
 Existing measurements put  bounds on sub-GeV   dark sector scenarios, with  $\Lambda$  up to $O(10-100)$\,MeV and $m$ up to  $\sim 3-50$\,MeV.
These DM force searches are  very complementary to nucleon-based direct detection. We have presented predictive cosmological scenarios which are constrained by these searches.


 \section*{Acknowledgements}
 I thank  E. Bertuzzo,  G. von Gersdorff, F. Goertz, G. Grilli de Cortona, F. Iocco, E. Pont\'on, R. Rosenfeld,  V. Sanz, and especially N. Bernal and C. S. Fong  for useful discussions. I  acknowledge W. G. Ubachs and  G. Pignol  for providing crucial clarifications on experimental aspects, and V. Korobov for important clarifications on molecular wave functions. This work is supported by the S\~ao Paulo Research Foundation (FAPESP) under grants \#2011/11973 and \#2014/21477-2.

\section*{References}

\bibliography{biblio}

\end{document}


\vspace*{5mm}

\begin{center}
\huge Supplemental Material
\end{center}
\vspace{1cm}

\section{Bounds}

\subsubsection*{\textit{Molecular spectroscopy}:} 
\begin{table}[h!]
\resizebox{\columnwidth}{!}{
\begin{tabular}{|c|c|c|c|c|c|c|c|c|c|c|c|}
\cline{3-12}
\multicolumn{2}{c|}{•} & ${\cal O}^0_a$ & ${\cal O}^0_b$ & ${\cal O}^0_c$ & ${\cal O}^{\nicefrac{1}{2}}_a$ & ${\cal O}^{\nicefrac{1}{2}}_b$ & ${\cal O}^{\nicefrac{1}{2}}_c$ & ${\cal O}^1_a$ & ${\cal O}^1_b$ & ${\cal O}^1_c$ & ${\cal O}^1_d$   \\
 \hline
 $dd\mu^+$ &\begin{tabular}{@{}c@{}} $\Lambda_{\rm max}${\scriptsize [MeV]} \\  $m$ {\scriptsize [MeV]} \end{tabular}  &  
 \begin{tabular}{@{}c@{}} $197$  \\  $3.1$ \end{tabular}  & 
 \begin{tabular}{@{}c@{}} $16.8$  \\  $21$ \end{tabular}  &
 \begin{tabular}{@{}c@{}} $30.9$  \\  $39$ \end{tabular}  &
 \begin{tabular}{@{}c@{}} $26.4$  \\  $33$ \end{tabular}  &
 \begin{tabular}{@{}c@{}} $23.8$  \\  $30$ \end{tabular}  &
 \begin{tabular}{@{}c@{}} $23.8$  \\  $30$ \end{tabular}  &
 \begin{tabular}{@{}c@{}} $30.9$  \\  $39$ \end{tabular}  &
 \begin{tabular}{@{}c@{}} $30.0$  \\  $38$ \end{tabular} &
 \begin{tabular}{@{}c@{}} $43.7$  \\  $55$ \end{tabular} &
  \begin{tabular}{@{}c@{}} $43.7$  \\ $55$ \end{tabular}
 \\  \hline
   HD$ ^+ $ &\begin{tabular}{@{}c@{}} $\Lambda_{\rm max}${\scriptsize [keV]} \\  $m$ {\scriptsize [keV]} \end{tabular}  &  
   \begin{tabular}{@{}c@{}} $76758$  \\  $8.7$ \end{tabular}  & 
   \begin{tabular}{@{}c@{}} $422$  \\  $21.7$ \end{tabular}  &
   \begin{tabular}{@{}c@{}} $151$  \\  $38.3$ \end{tabular} & 
   \begin{tabular}{@{}c@{}} $661$  \\  $21.7$ \end{tabular}  &
   \begin{tabular}{@{}c@{}} $597$  \\  $25.6$ \end{tabular}  &
   \begin{tabular}{@{}c@{}} $597$  \\  $21.7$ \end{tabular}  &
   \begin{tabular}{@{}c@{}} $151$  \\  $40.1$ \end{tabular}  &
   \begin{tabular}{@{}c@{}} $751$  \\  $20.7$ \end{tabular}  &
   \begin{tabular}{@{}c@{}} $214$  \\  $38.9$ \end{tabular} &
      \begin{tabular}{@{}c@{}} $214$  \\  $36.5$ \end{tabular}
 \\  \hline
   H$_2$ &\begin{tabular}{@{}c@{}} $\Lambda_{\rm max}${\scriptsize [keV]} \\  $m$ {\scriptsize [keV]} \end{tabular}  &  
    \begin{tabular}{@{}c@{}} $14732$  \\  $12.7$ \end{tabular}  & 
    \begin{tabular}{@{}c@{}} $263$  \\  $26.3$ \end{tabular}  &
    \begin{tabular}{@{}c@{}} $116$  \\  $44.0$ \end{tabular} & 
    \begin{tabular}{@{}c@{}} $412$  \\  $26.3$ \end{tabular}  &
    \begin{tabular}{@{}c@{}} $372$  \\  $30.8$ \end{tabular}  &
    \begin{tabular}{@{}c@{}} $372$  \\  $26.3$ \end{tabular}  &
    \begin{tabular}{@{}c@{}} $116$  \\  $46.0$ \end{tabular}  &
    \begin{tabular}{@{}c@{}} $468$  \\  $30.6$ \end{tabular}  &
    \begin{tabular}{@{}c@{}} $164$  \\  $44.7$ \end{tabular} &
    \begin{tabular}{@{}c@{}} $164$  \\  $42.0$ \end{tabular}
 \\  \hline
\end{tabular}
}
\caption{\label{tab:molecules} 
Limits on $\Lambda,m$ for the DM operators of Eq.\,(5). For each measurement, first line gives the limit on $\Lambda\equiv \Lambda_{\rm max}$ assuming $m=0$, second line gives the limit  on $m$ assuming $\Lambda=\Lambda_{\rm max}/10 $.
}
\end{table} 

Regarding the sensitivity on $m$ assuming $\Lambda=\Lambda_{\rm max}/10 $ in the case of the $dd\mu^+$ observable,  the value of $m$ often reaches the limit of validity of the EFT, which is taken here as $m\sim 4\pi \Lambda$. When this happens, this value for $m$ is reported in the table.

\subsubsection*{\textit{Neutron scattering}:}
\begin{table}[h!]
\begin{tabular}{|c|c|c|c|c|c|c|c|c|c|c|}
\cline{2-11}
\multicolumn{1}{c|}{•} & ${\cal O}^0_a$ & ${\cal O}^0_b$ & ${\cal O}^0_c$  & ${\cal O}^{\nicefrac{1}{2}}_a$ & ${\cal O}^{\nicefrac{1}{2}}_b$ & ${\cal O}^{\nicefrac{1}{2}}_c$ & ${\cal O}^{1}_a$ & ${\cal O}^{1}_a$ & ${\cal O}^{1}_c$ & ${\cal O}^{1}_d$ \\
 \hline
$l_{\rm tot}/l_{\rm opt}$ & 447 & 16 & 3.7 & 25 & 23 & 22 & 3.7 & 29 & 5.2 & 5.2 \\ \hline
$l_{\rm Bragg}/l_{\rm opt}$ & 13 & 1.3 & 0.6 & 2.1 & 0.2 & 1.9 & 0.6 & 1.9 & 0.9 & 0.2 \\ \hline \hline
$l_{\rm NP}^{\rm loc}$ & 2200
 & 16  & 3.0 & 25 & 0 & 22 & 3.0 & 22 & 4.5 & 0 \\ \hline
\end{tabular}
\caption{\label{tab:neutrons} 
Lower bounds  on $\Lambda$ [MeV] for the DM operators of Eq.\,(5) from neutron scattering,  assuming $m=100$\,keV for the DM mass. }
\end{table}

\begin{figure}[h!]
\begin{picture}(250,278)
\put(0,0){\includegraphics[width=9 cm,trim={0 0 0 0},clip]{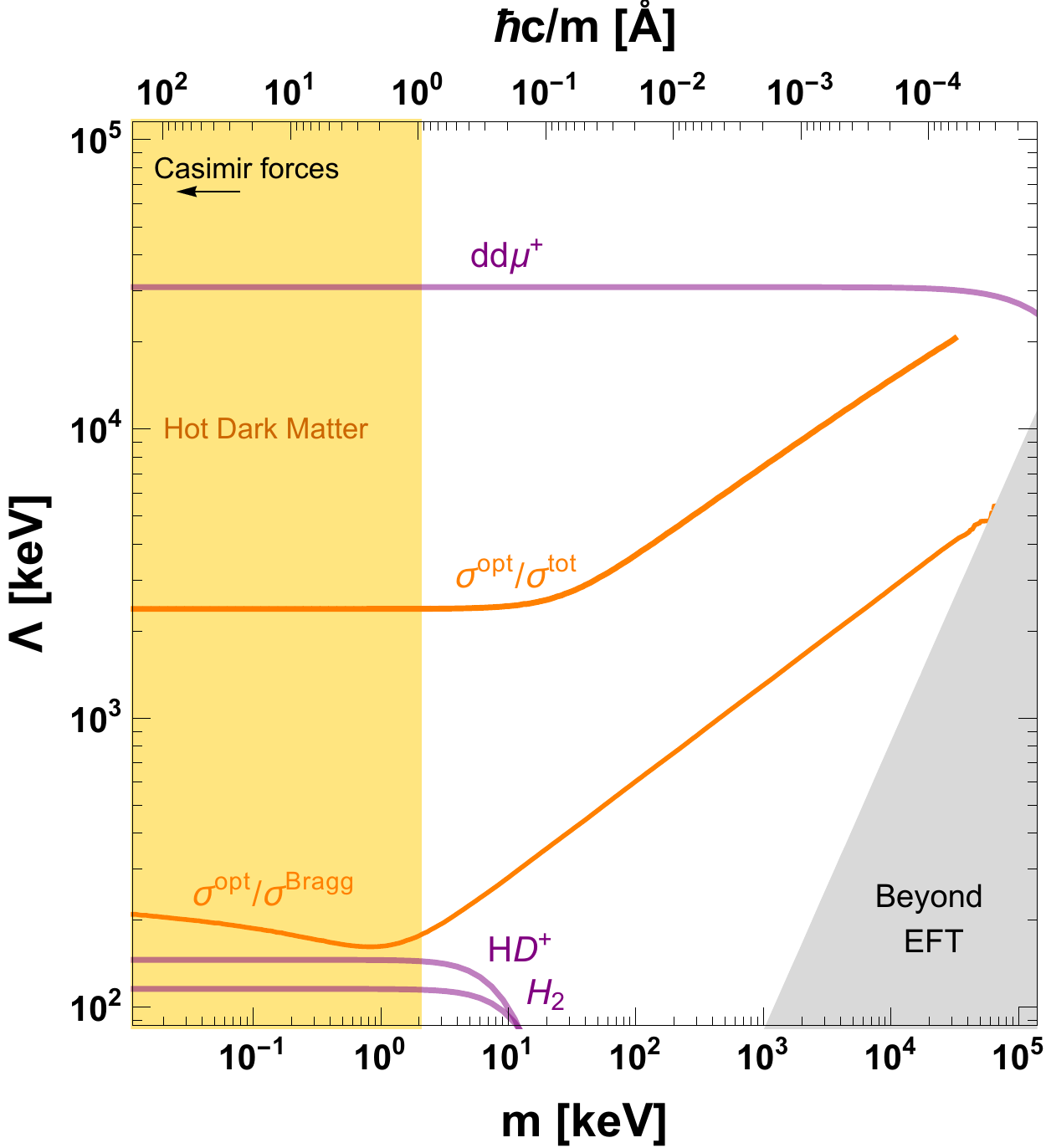}}
\put(141,225){\framebox{\Longstack[l]{ Complex scalar \\ $\bar N N  |\partial_\mu\phi|^2$ interaction}}}
\end{picture}
\caption{
Limits on the ${\cal O}^{0}_c$ (and approximately ${\cal O}^{1}_a$) interaction in the $m-\Lambda$ plane. Same conventions as in Fig.\,2. 
\label{fig:exclusions2}
}
\end{figure}

\section{Estimation of the annihilation cross sections }

\begin{figure}[h!]
\center
\includegraphics[width=6 cm,trim={0 0 0 0},clip]{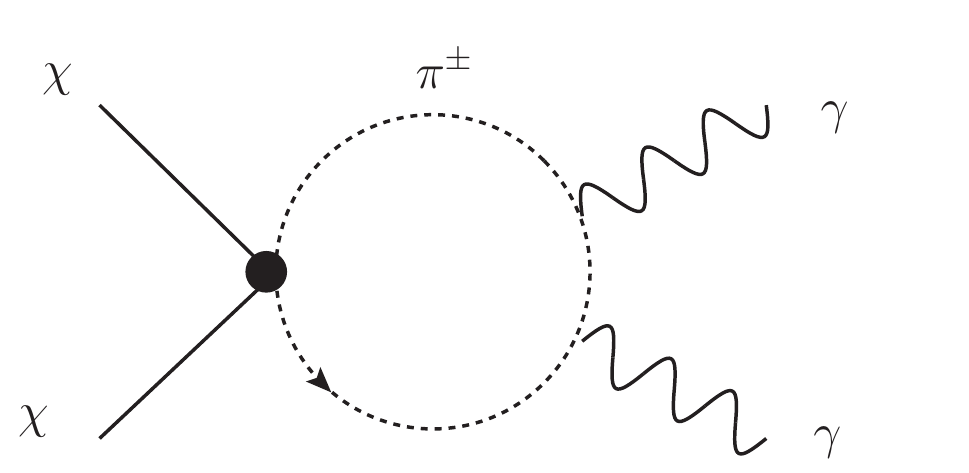}
\hspace{1cm}
\includegraphics[width=6 cm,trim={0 0 0 0},clip]{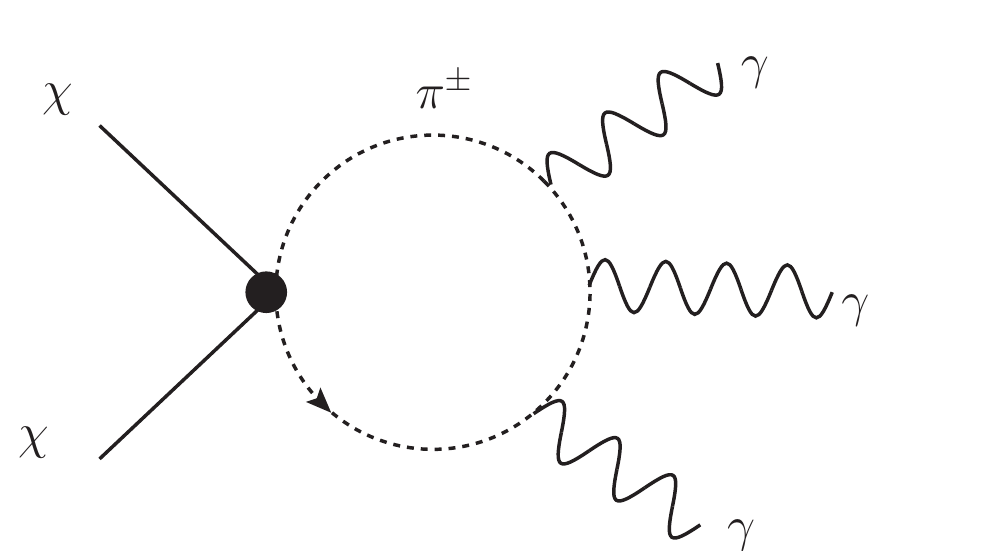}
\caption{ Diagrams of the pion-driven annihilations. The effective coupling is either a ``scalar channel'' (left) or a ``vector channel'' (right).
\label{fig:DM_graphs}
}
\end{figure}

Here are some details regarding the evaluation of annihilation cross sections via  pion loop given in Eq.~3. For concreteness we assume Dirac DM. The non-relativistic annihilation  cross-sections are estimated as follows. 
The effective couplings of DM to pions are assumed to be 
\be\frac{m_\pi}{\Lambda'^2 }\chi \chi \, \pi^+ \pi^- \,, \quad
\frac{1}{\Lambda'^2 }\chi \gamma^\mu \chi \,i \pi^+ \overleftrightarrow \partial_\mu \pi^- 
 \ee for the scalar and vector channels respectively. The coupling to pions is assumed to be of same order as the coupling to nucleons, hence one has $\Lambda'\sim\Lambda$.
 
For the scalar channel, the $\chi \chi \rightarrow \gamma \gamma$ amplitude is roughly estimated by
\be
{\cal M} \sim \frac{4\, s^3}{\Lambda^2\,m_\pi}\frac{e^2}{16\pi^2}\,.
\ee
The nonrelativistic annihilation cross section is then given by 
\begin{align}
\sigma_{2\gamma}=\frac{1}{16\pi^2 s} \frac{1}{4} \overline{|{\cal M}|}^2 \sim 
\frac{\alpha^2_{\rm em}m^4}{4 \pi^3 m_\pi^2 \Lambda^4}
\end{align}
where $s\approx 2m$.

For the vector channel, annihilation into $2\gamma$ is forbidden by Furry's theorem and the leading process is annihilation into 3 photons. To estimate the amplitude one uses the EFT  result for the $Z \gamma\gamma\gamma$ interaction induced by the loop of  a heavy charged scalar (see Ref.~[88]). We have
\be
{\cal M} \sim \frac{6\,s^6}{50 m_\pi^4\, \Lambda^2}\frac{e^4}{16\pi^2}
\ee
where the $~1/50$ is an approximate factor coming from the effective coupling induced by the pion loop [88]. To estimate the $2\rightarrow 3$ cross section we simply include an extra phase space factor $1/4\pi$. The $3\gamma$ cross section is estimated to be
\be
\sigma_{3\gamma} \sim  
\frac{288\, \alpha^3_{\rm em}\,m^{10}}{625\, \pi^2\, m_\pi^8\, \Lambda^4} \,.
\ee